\documentstyle[12pt]{article}

\def\lp{\left(}
\def\rp{\right)}
\def\lb{\left[}
\def\rb{\right]}
\def\ai{\'{\i}}
\def\pp{\varphi}
\def\be{\begin{equation}}
\def\ee{\end{equation}}

\begin{document}

\baselineskip.29in


\title{\bf{A note about the perturbative dynamics of symmetric shells}}

\author{Emilio Rub\'{\i}n de Celis$^{1,2,}$\thanks{e-mail: erdec@df.uba.ar} $$ and
Claudio Simeone$^{1,2,}$\thanks{e-mail: csimeone@df.uba.ar}\\ \\
{\small $^1$ Universidad de Buenos Aires. Facultad de Ciencias Exactas y Naturales. Departamento de F\'{\i}sica.} \\ 
{\small Buenos Aires, Argentina.}\\
{\small $^2$ CONICET - Universidad de Buenos Aires. Instituto de F\'{\i}sica de Buenos Aires (IFIBA).} \\ 
{\small Buenos Aires, Argentina.}}

\date{}

\maketitle







\noindent ABSTRACT: Several examples suggest the conjecture that the central aspect determining a monotonic evolution for perturbed highly symmetric thin-shells is the approximation adopted for their equations of state.

\vskip2cm


\vskip1cm


\newpage

\section{Introduction}

The characterization and dynamics of thin-shells has received considerable attention in the recent time, as they provide useful models for objects of astrophysical scale and also appear in the treatment of certain modern cosmological scenarios. Shells have also been proposed as a simplified description for the matter supporting traversable wormholes. Within this program, it was recently found that, under certain assumptions regarding their equations of state, the shells at the throat of cylindrically symmetric wormholes connecting two identical spacetimes would develop a monotonic evolution after a symmetric mechanical perturbation \cite{nos1,nos2,nos3}. Let us begin by  briefly reviewing how this result was obtained.  The most general case studied \cite{nos3} is that of a wormhole throat that connects two identical generic static spacetimes with metric 
\be
ds_{\pm}^2=-A(r)dt^2+B(r)dr^2+C(r)d\pp^2+D(r)dz^2.
\ee
The energy-momentum tensor of the matter shell placed at the throat surface $r=a$ is given by the following surface energy density $\sigma$ and pressures $p_\pp$ and $p_z$: 
\be
\sigma=- \frac{\sqrt{1 + B(a) \dot{a}^2}}{8 \pi \sqrt{B(a)}} \left[\frac{C'(a)}{C(a)} + \frac{D'(a)}{D(a)} \right],\label{e8}
\ee
\be
p_\pp=\frac{\sqrt{B(a)}}{8 \pi  \sqrt{1 + B(a)\dot{a}^2}} 
\left\{ 2 \ddot{a} +  \left[ \frac{D'(a)}{D(a)} + \frac{A'(a)}{A(a)} +\frac{ B'(a)}{B(a)} \right] \dot{a}^2 + \frac{1}{B(a)}\lb\frac{A'(a)}{A(a)} + \frac{D'(a)}{D(a)} \rb \right\},\label{e9}
\ee
\be
p_z=\frac{\sqrt{B(a)}}{8 \pi  \sqrt{1 + B(a) \dot{a}^2}}
\left\{ 2 \ddot{a} + \left[ \frac{C'(a)}{C(a)} + \frac{A'(a)}{A(a)} + \frac{B'(a)} {B(a)} \right] \dot{a}^2 +  \frac{1}{B(a)}\lb\frac{A'(a)}{A(a)} + \frac{C'(a)}{C(a)} \rb\right\},\label{e10}
\ee
where primes stand for derivatives with respect to the radial coordinate and dots stand for derivatives with respect to the proper time measured on the shell. The energy density and pressures of a static configuration, i.e. that corresponding to a static shell at $r=a_0$, are 
\be
\sigma _0= - \frac{1}{8 \pi \sqrt{B(a_0)}} \left[ \frac{C'(a_0)}{C(a_0)} + \frac{D'(a_0)}{D(a_0)} \right],\label{e12}
\ee
\be
p_{\varphi 0} =  \frac{1}{8 \pi \sqrt{B(a_0)}} \left[ \frac{A'(a_0)}{A(a_0)} +   \frac{D'(a_0)}{D(a_0)} \right],\label{e13}
\ee
\be
p_{z 0}=  \frac{1}{8 \pi \sqrt{B(a_0)}} \left[\frac{A'(a_0)}{A(a_0)} + \frac{C'(a_0)}{C(a_0)} \right] .\label{e14}
\ee
It is easy to check that the static energy density and pressures satisfy the relations
\be
p_{\varphi 0}=X_{\varphi 0}\sigma_0,\ \ \ \ \ \ \ \ X_{\varphi 0}\equiv -\frac{C(a_0)[A(a_0)D'(a_0)+A'(a_0)D(a_0)]}{A(a_0)[C(a_0)D'(a_0)+C'(a_0)D(a_0)]},\label{e15}
\ee
\be
p_{z 0}=X_{z0}\sigma_0,\ \ \ \ \ \ \ \ X_{z0}\equiv -\frac{D(a_0)[A(a_0)C'(a_0)+A'(a_0)C(a_0)]}{A(a_0)[C(a_0)D'(a_0)+C'(a_0)D(a_0)]}.\label{e16}
\ee
In \cite{nos3} analogous relations were assumed to be a valid approximation for the energy density and pressures corresponding to a shell undergoing a slow symmetric perturbation ($\dot a \ll 1$)\footnote{We adopt the usual units such that $G=c=1$.}. Consequently, the  general relations of the form $p_\varphi=Y_\varphi\sigma$, $p_z=Y_z\sigma$ which would follow from Eqs. (\ref{e8}), (\ref{e9}) and (\ref{e10}) were approximated by $p_\varphi=X_\varphi\sigma$, $p_z=X_z\sigma$, where $X_\varphi$ and $X_z$ are defined by substituting $a$ instead of $a_0$ in Eqs. (\ref{e15}) and (\ref{e16}). The result was the equation of motion for the shell 
\be
2B(a)\ddot a+B'(a){\dot a}^2=0, 
\ee
which gives the result of a monotonic evolution: the velocity of the shell (or, more precisely, the proper time derivative  of the radial coordinate of the shell placed at the wormhole throat) is given by
\be
\dot a(\tau)=\dot a_0\sqrt{\frac{B(a_0)}{B(a(\tau))}}.
\ee
The velocity can only increase or decrease its magnitude depending on the behaviour of the metric function $B(r)$, but cannot change its sign. Thus this analysis of the perturbative dynamics of such generic wormholes would show that no stable configurations are possible under the hypothesis adopted, in the sense that there would be no turning points for the motion after a perturbation given by a small  initial speed ($\dot a_0\ll 1$).

A particular example of this result had already been found \cite{nos1} for the gauge cosmic string symmetric wormhole which connects two identical metrics of the form \cite{vil}
\be
ds_{\pm}^2=-dt^2+dr^2+W^2r^2d\pp^2+dz^2.
\ee
The parameter $W$ ($0<W\leq 1$) describes the conicity of the geometry, the case $W=1$ corresponding to no angle deficit\footnote{The circumference length is $2\pi r W$, so that the angle deficit is $2\pi (1-W)$.}. In this simple case the  components of the shell energy-momentum tensor take the form
\be
\sigma=-\frac{\sqrt{1+{\dot a}^2}}{4\pi a},
\ee
\be
p_\pp=\frac{\ddot a}{4\pi \sqrt{1+{\dot a}^2}},
\ee
\be
p_z=\frac{1+{\dot a}^2+a\ddot a}{4\pi a \sqrt{1+{\dot a}^2}}.
\ee
In the static case we have
\be
\sigma_0=-\frac{1}{4\pi a_0}, \ \ \ \ \ \ p_{\pp 0}=0, \ \ \ \ \ \ p_{z0}=\frac{1}{4\pi a_0}.
\ee
It is easy to see that if the relation existing between the static energy density and pressures is adopted also for the case of a moving shell, the resulting vanishing angular pressure immediately forces the condition  $\ddot a =0$; then the shell evolution is not only monotonic but, moreover, it is uniform: $\dot a(\tau)=\dot a_0$.

Now, quite different behaviours of similar shells were obtained \cite{other} when other approximations were adopted for the equations of state. In particular, linear relations between the energy density and pressures led to the possibility of mechanical stability; in that approach, regions in parameter space corresponding to a positive second derivative of the ``potential'' $V(a)$ in an equation of motion of the form
\be
{\dot a}^2+V(a)=0
\ee 
were associated to stable static configurations about which the shell would oscillate if slightly perturbed\footnote{However, the interpretation of that kind of analysis is not devoid of certain subtleties; see, for instance, paragraph 4.1 of Ref. \cite{viswil}.}. The result of different evolutions associated to different equations of state is not surprising, and it certainly does not imply any flaw, but the natural question arises about which is the main aspect, if there is one, that determines a given type of evolution. In what follows, we study several examples which would suggest the conjecture that a monotonic evolution is mainly dictated by the approximation made in the choice of the equations of state.

\section{Simple examples} 

\subsection{Asymmetric gauge string wormhole} 

The same treatment of the simple example above can be extended to a gauge cosmic string {\it asymmetric} wormhole, which would connect two different conical spacetimes with metrics
\be
ds_{\pm}^2=-dt^2+dr_\pm^2+W_{\pm}^2r_\pm^2d\pp^2+dz^2.
\ee
In terms of the coordinates $x^\mu_- $, the wormhole throat is located at $r_-=a$, and in terms of the $x^\mu_+$, it is placed at $r_+=b$. The continuity of the metric across the shell placed at the throat implies that $a$ and $b$ must fulfil
\be
W_-^2a^2=W_+^2b^2.  
\ee
The application of the Darmois--Israel formalism \cite{daris} gives the relation between the geometries joined and the properties of the shell, more precisely the relation between the jump of the extrinsic curvature $[{K_i}^j]$ across the shell and the surface energy-momentum tensor ${S_i}^j={\rm diag}(-\sigma,p_\pp,p_z)$ of the matter on it: 
\be
8\pi {S_i}^j= -[{K_i}^j]+{\delta_i}^jK,
\ee 
where $K$ stands for the trace of $[{K_i}^j]$. In this simple example, and if we define $k=W_-/W_+$, the formalism gives the following expressions for the components of ${S_i}^j$:
\be
\sigma=-\frac{\sqrt{1+{\dot a}^2}}{8\pi a}-\frac{\sqrt{1+k^2{\dot a}^2}}{8\pi ka},
\ee
\be
p_\pp=\frac{\ddot a}{8\pi \sqrt{1+{\dot a}^2}}+\frac{k\ddot a}{8\pi  \sqrt{1+k^2{\dot a}^2}},
\ee
\be
p_z=\frac{\sqrt{1+{\dot a}^2}}{8\pi a}+\frac{\sqrt{1+k^2{\dot a}^2}}{8\pi ka}+\frac{\ddot a}{8\pi \sqrt{1+{\dot a}^2}}+\frac{k\ddot a}{8\pi \sqrt{1+k^2{\dot a}^2}}.
\ee
We see that the components of the energy-momentum tensor satisfy the relation $p_z=-\sigma+p_\pp$. The energy density and pressures for a static shell are 
\be
\sigma_0=-\frac{k+1}{8\pi ka_0},\ \ \ \ \ \ 
p_{\pp 0}=0,\ \ \ \ \ \
p_{z0}=\frac{k+1}{8\pi ka_0}.
\ee
If we assume that keeping the relations for the static case in the perturbative dynamic case is a reasonable approximation, we have $p_\pp=0$, and consequently:
\be
\ddot a =0,\ \ \  \ \ \ \  \dot a(\tau)=\dot a_0.
\ee
Hence, in this slight variation of the already studied case we again obtain a uniform motion of the shell placed at the wormhole throat.

\subsection{Shell joining two conical submanifolds}

If we are tempted to think that the monotonic evolution of the shell is a result of the non trivial topology of the examples considered (i.e. the existence of a throat which is a minimal area surface), a possible example to explore within simple cylindrically symmetric configurations is a spacetime with an inner region with a given deficit angle and an outer region with a different deficit angle:
\be
ds_{1,2}^2=-dt^2+dr_{1,2}^2+W_{1,2}^2r_{1,2}^2d\pp^2+dz^2.
\ee
The regions are joined by a shell which for the inner coordinate set is placed at  $r_1=a$, and for the outer one is placed at $r_2=b$.  The continuity of the metric across the joining surface dictates the relation
\be
W_1^2a^2=W_2^2b^2.
\ee
Introducing the definition  $k=W_1/W_2$ and aplying the thin-shell formalism as above, the components of the energy-momentum tensor for the thin layer at the surface $r=a$ read
\be
\sigma=\frac{\sqrt{1+{\dot a}^2}}{8\pi a}-\frac{\sqrt{1+k^2{\dot a}^2}}{8\pi ka},
\ee
\be
p_\pp=-\frac{\ddot a}{8\pi \sqrt{1+{\dot a}^2}}+\frac{k\ddot a}{8\pi \sqrt{1+k^2{\dot a}^2}},
\ee
\be
p_z=-\frac{\sqrt{1+{\dot a}^2}}{8\pi a}+\frac{\sqrt{1+k^2{\dot a}^2}}{8\pi ka}-\frac{\ddot a}{8\pi \sqrt{1+{\dot a}^2}}+\frac{k\ddot a}{8\pi  \sqrt{1+k^2{\dot a}^2}}.
\ee
As in the preceding example, we obtain the relation $p_z=-\sigma+p_\pp$.  In a static situation the energy density and pressures are 
\be
\sigma_0=\frac{k-1}{8\pi ka_0},\ \ \ \ \ \ p_{\pp 0}=0,\ \ \ \ \ \ 
 p_{z0}=\frac{1-k}{8\pi ka_0}.
\ee
We see that the static shell is constituted of normal $(\sigma_0>0)$ or exotic $(\sigma_0<0)$ matter depending on which submanifold presents a larger angle deficit. We also note that this feature is not changed by the shell motion. If, as before, we assume that the equations of state of the static case are a good approximation in a  perturbative evolution of the shell, then we have $p_\pp=0$ and consequently
\be
\ddot a =0,\ \ \  \ \ \ \  \dot a(\tau)=\dot a_0,
\ee
so the shell joining two conical submanifolds would also evolve with constant speed. Note that within this aproach the type of matter, reflected in the sign of the energy density, has no effect on the resulting kind of motion.

\section{More general cases}

\subsection{Wormholes in 2+1 dimensions}

One may wonder whether the monotonic motion of the shell is a peculiarity associated to pasting cylindrical submanifolds. Then let us examine a quite general $2+1$ example: wormholes connecting two equal $2+1-$dimensional symmetric spacetimes with metric
\be
ds_\pm^2=-f(r)dt^2+g(r)dr^2+h(r)d\theta^2.
\ee
The mathematical construction of a wormhole geometry with throat radius $a$ removes the regions $r<a$, and we assume that any finite radius associated to a zero or an infinity of the metric coefficients lies within that region. Then this analysis includes as a particular case, for example, the  wormhole which connects two equal geometries which are locally identical to the exterior (i.e. the region outside the horizon) of the well known Ba\~nados--Teitelboim--Zanelli (BTZ) black hole geometry \cite{btz}. The application of the Darmois--Israel formalism in the case of $2+1-$dimensional ``spherical'' symmetry and with symmetry across the throat gives the following linear energy density $\lambda$ and pressure $p$ for the shell placed at $r=a$:
\be
\lambda=-\frac{h'(a)}{8\pi h(a)}\sqrt{\frac{1+g(a){\dot a}^2}{g(a)}},\label{ll}
\ee
\be
p=\frac{1}{8\pi}\sqrt{\frac{g(a)}{1+g(a){\dot a}^2}}\left\{2\ddot a+{\dot a}^2\lb\frac{f'(a)}{f(a)}+\frac{g'(a)}{g(a)}\rb+\frac{f'(a)}{f(a)g(a)}\right\}.\label{pp}
\ee
In Ref. \cite{2+1} the approximation of a linearized equation of state was adopted for the matter on the shell; depending on the static radius $a_0$ and of the values of the parameters characterizing each class of geometry, this led to the possibility of stable configurations. Now let us consider the approximation that consists in assuming that the relation between $\lambda$ and $p$ valid for the static case holds after a slow perturbation. The components of the energy momentum tensor for a  static configuration read
\be
\lambda_0=-\frac{h'(a_0)}{8\pi h(a_0)\sqrt{g(a_0)}},
\ee
\be
p_0=\frac{f'(a_0)}{8\pi f(a_0)g(a_0)},
\ee
so that in this case the energy density and pressure are related by
\be
p_0=X_0\lambda_0, \ \ \ \ \ \ \ \ X_0\equiv -\frac{f'(a_0)h(a_0)}{f(a_0)h'(a_0)}.
\ee
If we assume an analogous relation for the general case, we have
\be
p=X\lambda, \ \ \ \  \ \ \ \ X\equiv -\frac{f'(a)h(a)}{f(a)h'(a)}.
\ee
Therefore, recalling (\ref{ll}) and (\ref{pp}), we obtain the condition
\be
\frac{f'(a)}{f(a)}\sqrt{\frac{1+g(a){\dot a}^2}{g(a)}}=\sqrt{\frac{g(a)}{1+g(a){\dot a}^2}}\left\{2\ddot a+{\dot a}^2\lb\frac{f'(a)}{f(a)}+\frac{g'(a)}{g(a)}\rb+\frac{f'(a)}{f(a)g(a)}\right\}.
\ee
This leads to the equation of motion 
\be
2\ddot a+{\dot a}^2 \frac{g'(a)}{g(a)}=0,
\ee
which has the solution
\be
\dot a(\tau)=\dot a_0\sqrt{\frac{g(a_0)}{g(a(\tau))}}.
\ee
Once again, there is no possibility of an oscillatory motion. The evolution is such that the shell speed cannot vanish, but it would increase or decrease with time depending on the behaviour of the metric function $g(r)$\footnote{Note that this has no relation with the flare-out condition required for the existence of a wormhole throat. The geodesics must open up at the throat, which in this case implies that the function $h$ must increase with the radial coordinate.}. An increasing velocity could lead to a problem with the validity of a perturbative treatment. Consider, for example, the exterior metric of the non charged and non rotating BTZ black hole at both sides of the wormhole throat: then $h(r)=r^2$ and $g(r)=f^{-1}(r)=(-M-\Lambda r)^{-1}$, with $\Lambda$ the cosmological constant, which must be negative to have a metric with the right signature. The function $g(r)$ increases towards the horizon radius $r_H=-M/\Lambda$ of the original manifolds and vanishes as $r\to \infty$. Hence an inwards perturbation would slow down, but an outwards perturbation would speed up with no limit, thus making eventually invalid the perturbative treatment.

\subsection{A class of $3+1-$dimensional  spherical wormholes}

A reasonable doubt about the example in $2+1$ dimensions could be posed, however, as it can be seen as a dimensional reduction of a cylindrical problem. Indeed, the analogy is apparent if we recall the energy-momentum tensor for a generic cylindrical thin-shell wormhole. Then, let us consider an essentially different case: a class of thin-shell wormholes connecting two identical quite generic $3+1-$dimensional spherically symmetric spacetimes. The metric at each side of the wormhole throat placed at $r=a$ is
\be
ds_\pm^2=-f(r)dt^2+f^{-1}(r)dr^2+r^2\lp d\theta^2+\sin^2\theta d\varphi^2\rp.
\ee
As before, we assume that any zero or infinity of the metric coefficients lying at finite radii is confined  within the removed regions $r<a$. Hence this includes the well-known wormholes connecting, among others, the outer parts of Schwarzschild and Reissner--Nordstr\"om geometries \cite{sphwh}. The application of the thin-shell formalism gives the following energy density and pressure for the shell placed at the wormhole throat:
\be
\sigma=-\frac{1}{2\pi a}\sqrt{f(a)+{\dot a}^2},
\ee
\be
p=\frac{2a\ddot a+2{\dot a}^2+2f(a)+af'(a)}{8\pi a\sqrt{f(a)+{\dot a}^2}}.
\ee
In \cite{sphwh} an equation of state of the form $p=p_0+\eta(\sigma-\sigma_0)$ was the model adopted for the shell matter.  Then the   condition $V''(a_0)>0$ on the potential of the equation of motion ${\dot a}^2+V(a)=0$ corresponded to the possibility of stable static configurations. Here, instead, we follow the approach proposed in Refs. \cite{nos1,nos2,nos3}. The pressure can be put in the form
\be
p=-\frac{\sigma}{2}+\frac{2\ddot a+f'(a)}{8\pi\sqrt{f(a)+{\dot a}^2}};
\ee
then, the static energy density and pressure are given by 
\be
\sigma_0=-\frac{\sqrt{f(a_0)}}{2\pi a_0},
\ee
\be
p_0=-\frac{\sigma_0}{2}+\frac{f'(a_0)}{8\pi \sqrt{f(a_0)}}.
\ee
It can be easily verified that the relation
\be
p_0=X_0\sigma_0, \ \ \ \ \ X_0\equiv -\frac{1}{2}\lb 1+\frac{a_0f'(a_0)}{2f(a_0)}\rb
\ee
holds for the static configuration. On the other hand, the relation between the energy density and pressure in the general case of a moving shell reads
\be
p=Y\sigma, \ \ \ \ \ Y\equiv  -\frac{1}{2}\lb 1+\frac{2a\ddot a +a f'(a)}{2f(a)+2{\dot a}^2}\rb.
\ee
Thus, if we impose the approximation that the energy density and pressure fulfil the same kind of relation in a perturbative evolution as they verify in a static configuration, we must substitute the fixed radius $a_0$ by the variable radius $a$ in $X_0$, to obtain 
\be
p=X\sigma, \ \ \ \ \ X\equiv -\frac{1}{2}\lb 1+\frac{af'(a)}{2f(a)}\rb.
\ee
This approximation forces $Y=X$, which after a little algebra leads to
\be
f'(a){\dot a}^2=2f(a)\ddot a.
\ee
This equation of motion has the solution 
\be
\dot a (\tau)=\dot a_0 \sqrt{\frac{f(a(\tau))}{f(a_0)}}
\ee
for the shell speed; hence, again, only monotonic evolutions of the shell radius are possible within this approach. Of course, depending on the form of the function $f(r)$ the speed would have different behaviours after an initial perturbation.  For example, in the case of a symmetric wormhole connecting the exterior part of two Schwarzschild geometries (that is, $a$ is larger than the Schwarzschild radius $2M$), the function $f(r)=1-2M/r$ increases with $r$; hence if the shell is perturbed inwards, then $a(\tau)<a_0$ and the speed would decrease. In the case of an outwards perturbation, instead, we have $a(\tau)>a_0$ and the speed would grow, which could render the perturbative approach eventually invalid. More precisely: because in this case the function $f(r)$  has an upper limit ($f(r)\to 1$ as $r\to \infty$), a small positive initial speed ($0<\dot a_0\ll 1$) eventually leads to $\dot a\to {\dot a}_0/\sqrt{1-2M/a_0}$; hence a slow evolution is only ensured if the starting configuration is far away from the Schwarzschild radius.

\subsection{Spherical $3+1-$dimensional shells}

As the final case, we address the perturbative dynamics of a spherically symmetric shell separating an inner from an outer region. This problem was studied in the linearized approach in, for example,  Refs. \cite{shells} and also in \cite{dims} in other spacetime dimensions. This class of geometries avoids the possible pecularities associated with the equality of the metric functions at both sides of the shell, as it is the situation in wormholes symmetric across the throat, and, mainly, with the globally non trivial topology of manifolds with a minimal area surface. Then we consider two manifolds ${\cal M}_1$ and  ${\cal M}_2$ with metrics of the form
\be
ds_{1,2}^2=-f_{1,2}(r)dt_{1,2}^2+f_{1,2}^{-1}(r)dr^2+r^2\lp d\theta^2+\sin^2\theta d\varphi^2\rp,
\ee
and we study the dynamics of the shell at the surface $r=a$ joining the region $r<a$ of the first manifold with the region $r>a$ of the second one\footnote{If horizons exist in any submanifold, we take the shell radius larger than the outer horizon radius.}.  The continuity of the line element across the shell is achieved by suitably defining the time coordinate at each side; the Darmois--Israel formalism gives the following  components of the surface energy momentum tensor for the moving shell:
\be
\sigma= \frac{1}{4\pi a}\lb\sqrt{f_1(a)+{\dot a}^2}-\sqrt{f_2(a)+{\dot a}^2}\rb,\label{00}
\ee
\be
p=-\frac{\sigma}{2}+\frac{1}{16\pi}\lb\frac{2\ddot a+f'_2(a)}{\sqrt{f_2(a)+{\dot a}^2}}-\frac{2\ddot a+f'_1(a)}{\sqrt{f_1(a)+{\dot a}^2}}\rb.\label{pres}
\ee
In the case of a static shell these expressions reduce to
\be
\sigma_0= \frac{1}{4\pi a_0}\lb\sqrt{f_1(a_0)}-\sqrt{f_2(a_0)}\rb,\label{000}
\ee
\be
p_0=-\frac{\sigma_0}{2}+\frac{1}{16\pi}\lb\frac{f'_2(a_0)}{\sqrt{f_2(a_0)}}-\frac{f'_1(a_0)}{\sqrt{f_1(a_0)}}\rb.\label{pres0}
\ee
If we introduce the relations $p_0=X_0\sigma_0$ and $p=Y\sigma$ as done in the preceding examples, from Eqs. (\ref{00})--(\ref{pres0}) we have
\be
X_0=-\frac{1}{4}\lb 2+a_0\frac{f'_2(a_0)/\sqrt{f_2(a_0)}-f'_1(a_0)/\sqrt{f_1(a_0)}}{\sqrt{f_2(a_0)}-\sqrt{f_1(a_0)}}\rb,
\ee
\be
Y=-\frac{1}{4}\lb 2+a\frac{\lp f'_2(a)+2\ddot a\rp/\sqrt{f_2(a)+{\dot a}^2}-\lp f'_1(a)+2\ddot a\rp/\sqrt{f_1(a)+{\dot a}^2}}{\sqrt{f_2(a)+{\dot a}^2}-\sqrt{f_1(a)+{\dot a}^2}}\rb.
\ee
The assumption that the relation between the energy density and pressure in a slow evolution of the shell can be approximated by the relation verified in a static configuration introduces the equation $p=X\sigma$, where
\be
X=-\frac{1}{4}\lb 2+a\frac{f'_2(a)/\sqrt{f_2(a)}-f'_1(a)/\sqrt{f_1(a)}}{\sqrt{f_2(a)}-\sqrt{f_1(a)}}\rb,
\ee 
and the condition $X=Y$. From this, a first integration gives
\be
\frac{\sqrt{f_2(a)+\dot a^2}-\sqrt{f_1(a)+\dot a^2}}{\sqrt{f_2(a_0)+\dot a^2_0}-\sqrt{f_1(a_0)+\dot a^2_0}}=\frac{\sqrt{f_2(a)}-\sqrt{f_1(a)}}{\sqrt{f_2(a_0)}-\sqrt{f_1(a_0)}},
\ee
so that the speed at a given proper time $\tau$ and the initial speed are related in a rather cumbersome way. Indeed, a long but straightforward calculation leads to the result
\be
{\dot a}^2(\tau)=\frac{1}{4C_0^2}\left\{ C_0^4\lb\sqrt{f_2(a)}-\sqrt{f_1(a)}\rb^2-2C_0^2\lb f_2(a)+f_1(a)\rb+\lb\sqrt{f_2(a)}+\sqrt{f_1(a)}\rb^2\right\},
\ee
where $C_0=\lb\sqrt{f_2(a_0)+\dot a^2_0}-\sqrt{f_1(a_0)+\dot a^2_0}\rb/\lb\sqrt{f_2(a_0)}-\sqrt{f_1(a_0)}\rb$. Hence now, differing from the examples associated to wormholes symmetric across the throat, in a general case it is not possible to give the velocity of the shell as a simple function of the radius $a$ and the initial speed. However, for any functions $f_1$ and $f_2$ with given values of the parameters involved we can always treat the problem numerically. A physically interesting case is, clearly, the example of two Schwarzschild geometries associated to masses $M_1<M_2$, joined at the surface $r=a>2M_2$. This configuration implies a shell of normal matter beyond (i.e. outside) the largest of the two Schwarzschild radii. The numerical evaluation of this example is easily performed and gives, once again, a shell which after a perturbation given by a small initial speed evolves monotically, that is, it undergoes a motion without the possibility of turning points. Besides, the calculation shows that, as in the previous example, an inwards perturbation would slow down, and because in this case both $f_1\to 1$ and $f_2\to 1$ as $r\to \infty$, after an outwards perturbation the speed would grow but it would stay bounded. Moreover, for any $f_1$ and $f_2$ with such asymptotic behaviour we have that ${\dot a}^2\to(1-C_0^2)/C_0^2$ as $a\to\infty$. 
As we assume $\dot a^2_0\ll 1$, in the Schwarzschild case with the initial configuration $a_0 \gg 2M_2 > 2M_1$ we can easily check that $C_0^2$ is just slightly smaller than unity. Therefore, under these conditions, an outwards perturbation of the shell would be followed by a slow monotonic expansion.

{\section{Summary}

While it is to be expected that different equations of state relating the pressure(s) and energy density of shells undergoing a perturbative evolution lead to different behaviours, it is natural to wonder about which is the central aspect of a given problem determining a definite type of motion. This question is particularly interesting when quite different results were obtained in Refs. \cite{nos1,nos2,nos3} and in Refs. \cite{other} for similar problems. Here we have applied the approach adopted in Refs. \cite{nos1,nos2,nos3} to several different examples, and the result in all the cases considered --which include rather generic configurations-- is the same kind of evolution obtained in \cite{nos1,nos2,nos3}. This would suggest the conjecture that the resulting monotonic evolution is mainly determined by assuming that the form of the relation(s) between the energy density and pressure(s) of a static shell can be considered a good approximation for the same shell undergoing a symmetric perturbation.

\section*{Acknowledgments}

This work has been supported by Universidad de Buenos Aires and CONICET.

\end{document}